# Estimation of migrate histories of the Japanese sardine in the Sea of Japan by combining the microscale stable isotope analysis of otoliths and a data assimilation model

**Tomoya Aono[1, 2*], Tatsuya Sakamoto[3**], Toyoho Ishimura[2, 4*], Motomitsu Takahashi[3], Tohya Yasuda[5], Satoshi Kitajima[3], Kozue Nishida[2, 6], Takayoshi Matsuura[2], Akito Ikari[2], Shin-ichi Ito[1*]**

[1]Atmosphere and Ocean Research Institute, The University of Tokyo, Kashiwa, Chiba, Japan

[2]National Institute of Technology, Ibaraki College, Hitachinaka, Ibaraki, Japan

[3]Fisheries Resources Institute, Japan Fisheries Research and Education Agency, Nagasaki, Japan

[4]Graduate School of Human and Environmental Studies, Kyoto University, Kyoto, Japan

[5]Fisheries Resources Institute, Japan Fisheries Research and Education Agency, Yokohama, Kanagawa, Japan

[6]Graduate School of Life and Environmental Sciences, University of Tsukuba, Tsukuba, Ibaraki, Japan

**\* Correspondence:**
Tomoya Aono
aontom09@gmail.com
Toyoho Ishimura
ishimura.toyoho.8r@kyoto-u.ac.jp
Shin-ichi Ito
goito@aori.u-tokyo.ac.jp

**\*\*Present Address:** Instituto Português do Mar e da Atmosfera (IPMA), Rua Alfredo Magalhães Ramalho, 6, 1495-006 Lisbon, Portugal

**Abstract**

The Japanese sardine (*Sardinops melanostictus*) is a small pelagic fish found in the Sea of Japan, the marginal sea of the western North Pacific. It is an important species for regional fisheries, but their transportation and migration patterns during early life stages remain unclear. In this study, we analyzed the stable oxygen isotope ratios of otoliths of young-of-the-year (age 0) Japanese sardines collected from the northern offshore and southern coastal areas of the Sea of Japan in 2015 and 2016. The ontogenetic shifts of the geographic distribution were estimated by comparing the profiles of life-long isotope ratios and temporally varying isoscape, which was calculated using the temperature and salinity fields produced by an ocean data assimilation model. Individuals that were collected in the northern and southern areas hatched and stayed in the southern areas (west offshore of Kyushu) until late June, and thereafter, they can be distinguished into two groups: one that migrated northward at shallow layer and one that stayed around the southern area in the deep layer. A comparison of somatic growth trajectories of the two groups, which was reconstructed based on otolith microstructure analysis, suggested that individuals that migrated northward had significantly larger body lengths in late June than those that stayed in the southern area. These results indicate that young-of-the-year Japanese sardines that hatched in the southern area may have been forced to choose one of two strategies to avoid extremely high water temperatures within seasonal and geographical limits. These include migrating northward or moving to deeper layers. Our results indicate that the environmental variabilities in the southern area could critically impact sardine population dynamics in the Sea of Japan.

# 1    Introduction

The Japanese sardine (*Sardinops melanostictus*) is a species with large biomass fluctuations. The abundance of Japanese sardines showed substantial fluctuations of several orders of magnitude during the last 3000 years (Kuwae et al., 2017), and fluctuations have also been reported regarding the associated fisheries catch in recent centuries (Yasuda et al., 2019). These fluctuations are considered to be driven by environmental variabilities, which are often represented by the Pacific Decadal Oscillation (e.g., Chavez et al., 2003). Clarification of the mechanistic links between climate and marine environments and population fluctuation is necessary for the accurate prediction of resource abundance and fisheries management. Variations in survival rates during the post-hatching larval and juvenile stages are hypothesized to be of great importance for population fluctuations (e.g., Hjort, 1914). Therefore, the knowledge of the geographical distributions and environment of Japanese sardines during its critical life stages are essential for revealing the impact of climate and marine environment changes on population fluctuations.

Current fisheries management of the Japanese sardine assumes two management units or stocks—the Pacific stock (distributed in the western North Pacific) and the Tsushima Warm Current Stock (distributed in the Sea of Japan), both of which show large fluctuations in abundance at similar time scales. While the patterns of transport and migration during the early life stages have been reported for the Pacific stock (e.g., Okunishi et al., 2009; Sakamoto et al., 2019), such knowledge for the Tsushima Warm Current Stock is insufficient. Spawning of the Tsushima Warm Current Stock occurs in coastal areas from west Kyushu to the Noto Peninsula from winter to early summer (January–June, Fig. 1). The main spawning ground changes depending on the sea surface temperature and spawning stock biomass (Furuichi et al., 2020). The young-of-the-year Japanese sardine is widely found from the off Noto Peninsula to the west coast of Kyushu in late summer (Ito, 1961; Yasuda et al., 2021). A recent study showed that the environment of the northern offshore area of the Sea of Japan with larger and lipid-rich prey zooplankton is suitable for the energy acquisition of sardine juveniles (Yasuda et al., 2021). In addition, the distribution of the stock is known to expand and shrink in response to population fluctuations; during periods of increased biomass, the distribution of the stock area expands to the northern offshore area of the Sea of Japan (Muko et al., 2018). These observations lead to the hypothesis that the successive and successful transportation (or migration) of juveniles to the northern offshore area might be the driver of the increase in stock (Muko et al., 2018; Yasuda et al., 2021). However, the origin, transportation, and migration routes of the individuals that reach the northern offshore area of the Sea of Japan have not yet been clarified. Understanding the movement patterns and their variety, and mechanisms that create the variety may, therefore, provide insights into energy acquisition strategies, population dynamics, and ultimately management of sardines in the region.

The chemical composition of otoliths is a promising tool for revealing transportation and migration patterns during early life stages (e.g., Sturrock et al., 2012; Mu et al., 2021). Otolith is a hard tissue composed of calcium carbonate ($CaCO_3$: aragonite) that continuously grows every day, forming one growth ring per day, which allows the inference of the hatch date (Campana, 1990). Once it crystallized, the calcium carbonate composition remained unchanged. Therefore, the environmental history of individuals is preserved in the otolith in terms of chemical composition (Campana, 1999). The stable oxygen isotope ratio ($\delta^{18}O$) is a chemical signal that is frequently used to trace the migration of marine organisms (Hanson et al., 2013; Darnaude et al., 2014; Shiao and Sui, 2016; Torniainen et al., 2017; Darnaude and Hunter, 2018; Sakamoto et al., 2019; Chiang et al., 2020). The $\delta^{18}O$ of the otolith ($\delta^{18}O_{oto}$) is mainly affected by two parameters: the ambient water temperature and $\delta^{18}O$ of the surrounding seawater ($\delta^{18}O_{water}$) (Kim et al., 2007). $\delta^{18}O_{water}$ is strongly correlated with salinity in region-specific linear relationships (e.g., LeGrande and Schmidt, 2006; Kodaira et al., 2016). This indicates that by comparing the $\delta^{18}O_{oto}$ actually analyzed and *the* $\delta^{18}O_{oto}$ isoscape predicted from temperature and salinity fields, the location of the fish can be inferred (Torniainen et al., 2017). Recent developments in high-precision micromilling systems (Sakai, 2009) and microscale isotopic analytical methods (Ishimura et al., 2004, 2008) have realized $\delta^{18}O_{oto}$ analyses with a temporal resolution of 10 days for the Japanese sardine (Sakamoto et al., 2019), which may allow reconstruction of early life migration histories of the Japanese sardine in the Sea of Japan.

In this study, we aimed to understand the seasonal movements of larval and juvenile Japanese sardines in the Sea of Japan by using $\delta^{18}O_{oto}$. High temporal resolution analysis of $\delta^{18}O_{oto}$ was conducted for the young-of-the-years collected in the northern offshore and southern coastal areas of the Sea of Japan during the late summer of 2015 and 2016. The relationship between $\delta^{18}O_{water}$ and salinity in the entire Sea of Japan from 2015 and 2016 was also established to allow the prediction of $\delta^{18}O_{water}$ from salinity. Using the relationship, temperature, and salinity outputs from an ocean data assimilation model, possible ontogenetic shifts in the distribution of the young-of-the-year Japanese sardines were estimated. As the existence of two different migratory types, namely the northward migration group and resident group, was indicated, we tested whether biological characteristics (body length and growth rate) were related to the selection of the migration patterns.

## 2    Materials & Methods

### 2.1    Fish sampling and Otolith $\delta^{18}O$ analysis

The young-of-the-years of Japanese sardines were collected for otolith analyses in three sampling areas (off Noto Peninsula, Tsushima Strait, and off Goto Islands) in the Sea of Japan from August to September in 2015 and 2016 by the R/V Yoko-maru, Japan Fisheries Research and Education Agency (Fig. 1; Table 1). Fish were frozen immediately after capture and preserved in -80 or -20 °C. Thereafter, the specimens were thawed in a laboratory on land and their scaled body length (BL) and body weight were measured. Sagittal otoliths were extracted, cleaned using a brush, rinsed with ethanol and distilled water, and dried at room temperature for several hours. The dried otoliths were embedded in epoxy resin (*p*-resin, Nichika Inc.) on a glass slide, and then placed in a dryer at 40 °C for at least a day to fix the otoliths. Thereafter, the embedded otoliths were polished with sandpaper (No. 1000, 2000) until the otolith core was exposed from the epoxy resin, and then the entire otolith was polished with alumina suspension (BAIKOWSKI International Corporation) to ensure that the daily rings could be easily observed under a microscope. The number of daily rings of the otolith and otolith daily increment widths were measured using an otolith measurement system (RATOC System Engineering Co. Ltd.). The otoliths of 154 individuals from 18 stations were used for daily ring measurements (Table 1).

The $\delta^{18}O_{oto}$ was analyzed for 17 of the 154 individuals. The otolith portions that were formed every 10–30 days were milled sequentially from the edge to the core using a high-precision micro-milling system (GEOMILL326, Izumo-Web Ltd., Japan) (an example of the milling procedure is shown in Supplementary Video 1). The $\delta^{18}O_{oto}$ was determined using an IsoPrime100 isotope ratio mass spectrometer (Isoprime Ltd., Cheadle Hulme, UK) equipped with a customized continuous-flow gas preparation system (MICAL3c) at the National Institute of Technology, Ibaraki College, Hitachinaka, Japan. This system is capable of analyzing $\delta^{18}O$ of calcium carbonates of > 0.2 $\mu$g with an analytical precision of within ± 0.1 ‰, which allows high sensitivity and accuracy analysis (Ishimura et al., 2004, 2008; Nishida and Ishimura, 2017). Otolith powders were reacted with phosphoric acid at 25 °C, and the evolved $CO_2$ was purified and introduced into the mass spectrometry system. The $\delta^{18}O$ values of each sample were reported in standard $\delta$ notation (‰) relative to the Vienna Pee Dee Belemnite (VPDB) standard. We used the calcite acid fractionation factor of 1.01025 (Sharma and Clayton, 1965) to calculate isotopic values in order to compare them with isotopic values reported in previous studies. The analytical precision was better than ± 0.1 ‰ for $\delta^{13}C$ and $\delta^{18}O$. Some samples were excluded from the results because of insufficient $CaCO_3$ or handling failure (Table S1).

## 2.2 Seawater sampling and relationship between seawater $\delta^{18}O$ and salinity

Seawater samples were collected from 237 stations in the Sea of Japan and East China Sea between July 27th and September 15th in 2015, and March 9th to September 18th in 2016 which was selected from a dataset of Kodama et al. (submitted). Samples were collected using a bucket from the surface of the sea or using a pump that was installed at the bottom of a ship (≤ 5 m). Salinity was measured using a salinometer (Autosal8400B, Guildline) for the seawater collected with the bucket, whereas it was measured using a thermosalinometer (SBE45, Seabird) for the seawater collected from the bottom of the ship. Samples for the $\delta^{18}O$ analysis were poured into two high density polyethylene bottles or glass bottles (the bottles were rinsed twice with the seawater before use). To avoid evaporation of water, the bottles were sealed tightly and stored at 5°C until further analysis was conducted on land. The $\delta^{18}O$ in the seawater samples ($\delta^{18}O_{water}$) was determined using a wavelength-scanned cavity ring-down spectroscopy isotopic water analyzer (L2130-i; Picarro Inc., Santa Clara, CA, USA) at the National Institute of Technology, Ibaraki College. All samples were subsampled and filtered through a 0.45 $\mu m$ filter prior to analysis. The long-term external precision was within ± 0.05 ‰ for $\delta^{18}O$. Isotopic values of water samples were reported relative to Vienna Standard Mean Ocean Water (VSMOW).

## 2.3    Estimation of migration history

The following equation relating $\delta^{18}O_{oto}$ (‰) and water temperature $T$ (°C) was established for Japanese sardines using laboratory rearing experiments (Sakamoto et al., 2017):

$$\delta^{18}O_{oto} - \delta^{18}O_{water} = -0.18 \times T + 2.69 \;\cdots(1)$$

Based on Equation (1), $\delta^{18}O_{oto}$ can be estimated if the temperature and $\delta^{18}O_{water}$ are given. The $\delta^{18}O_{water}$ and salinity $S$ during 2015 and 2016 in the Sea of Japan and East China Sea showed significant positive correlations, as represented by the following equation:

$$\delta^{18}O_{water} = a \times S + b \;\cdots(2)$$

From Equations (1) and (2), the $\delta^{18}O$ of otoliths that would be produced at each location ($\delta^{18}O_{est}$) can be expressed as

$$\delta^{18}O_{est} = (-0.18 \times T + 2.69) + (a \times S + b) \;\cdots(3)$$

Based on Equation (3), $\delta^{18}O_{est}$ can be estimated if the temperature and salinity are known. Thus, the isoscape of $\delta^{18}O_{est}$ can be drawn from datasets of horizontal distributions of temperature and salinity fields, and areas where $\delta^{18}O_{est}$ is similar to the analyzed $\delta^{18}O_{oto}$ can be considered as possible locations for the fish.

At the medians of the date ranges corresponding to each $\delta^{18}O_{oto}$, water temperature and salinity in the Sea of Japan and East China Sea (125–142 °E, 30–50 °N) were used to calculate the isoscape of $\delta^{18}O_{est}$. Temperature and salinity data were extracted from an ocean assimilation model, the FRA-ROMS (Kuroda et al., 2013, 2017). The FRA-ROMS ocean forecast system was developed by the Japan Fisheries Research and Education Agency (FRA) to conduct ocean forecasts using assimilated initial conditions. In the present study, we used the reanalysis data of FRA-ROMS, which is based on the Regional Ocean Modelling System (ROMS; Haidvogel et al., 2008) as an ocean circulation model with three-dimensional variational (3D-Var) analysis schemes as a data assimilation method. Satellite-derived sea surface temperature and sea surface height and in-situ hydrographic observations of temperature and salinity were assimilated. The horizontal resolution in the Sea of Japan is 1/10-degree, with the vertical structure comprising 48 layers defined by a specific S-coordinate function in both models. Further detailed model specifications can be found in Kuroda et al. (2013, 2017), and its accuracy assessments can be found in Sakamoto et al. (2019). Because the Japanese sardine has mainly been found in the surface mixed layer (e.g., Yatsu et al., 2005) and mixed layer depths are generally 10–50 m in this region from spring (March–May) to summer (June–August), we used temperature and salinity data at a depth of 10 m or 30 m. Equation (3) was used to calculate $\delta^{18}O_{est}$, and the areas at which the difference between $\delta^{18}O_{est}$ and $\delta^{18}O_{oto}$ was less than or equal to ± 0.18 ‰ were considered as estimated distribution areas of the Japanese sardine. Here, 0.18 ‰ corresponds to the root mean square error of Equation (1) (Sakamoto et al., 2017), which represents the sum of biological and analytical errors.

To validate the estimated results, the estimated distribution area on the date closest to the sampling date for each individual was compared with the actual sampling area. In addition, the estimated distribution area on the date closest to the hatch date was compared with the results of a monthly egg and larval survey conducted at prefectural fisheries field stations (Oozeki et al., 2007; Furuichi et al., 2020).

## 2.4 Statistical analysis

To examine the seasonal variation in $\delta^{18}O_{oto}$, we calculated the monthly mean of $\delta^{18}O_{oto}$ for all individuals and compared the monthly mean of $\delta^{18}O_{oto}$ in the preceding and following months using the Mann-Whitney U test (excluding February, March, and September, when individuals from both years were not included). To compare the $\delta^{18}O_{oto}$ history of individuals between sampling areas, the monthly mean $\delta^{18}O_{oto}$ of individuals in each sampling area was compared using the Kruskal-Wallis test for each month. When a significant difference was detected, a post-hoc Steel-Dwass test was performed to identify the sampling areas that had significant differences.

To examine the differences in biological characteristics between potentially different migratory types, namely the northward migration group (off Noto Peninsula individuals) and resident group (Tsushima Strait and off Goto Islands individuals; see Results and Discussion), we compared somatic growth trajectories between these groups. The Mann-Whitney U test for comparisons of hatching date and BL at the time of sampling was conducted between groups. We also compared the BL before sampling. BL before the time of sampling was calculated using the biological intercept method, assuming a linear relationship between the otolith radius and BL with fixed 5.9 mm of BL at the deposition of the first increment of the otolith, following Takahashi et al. (2008). To test for differences in the daily growth rates among the groups, anomalies of the 3-day running-mean otolith increment width normalized by standard deviation were compared among the three sampling areas using the Kruskal-Wallis test. The anomalies were calculated in two ways: for each day age of the fish and for each calendar day (days from January 1st each year). When a significant difference was detected, a post-hoc Steel-Dwass test was performed to identify areas with significant differences.

## 3    Results

### 3.1    Otolith $\delta^{18}O$ analysis

The average $\delta^{18}O_{oto}$ gradually decreased from +0.08 ± 0.27 ‰ at the core to –1.36 ± 0.24 ‰ at the edge in 2015 and from –0.07 ± 0.44 ‰ at the core to –1.41 ± 0.11 ‰ in 2016 (Fig. 2). The $\delta^{18}O_{oto}$ values of individuals in 2015 varied between +0.6 ‰ and –2.0 ‰, and those of individuals in 2016 varied between +0.5 ‰ and –2.2 ‰. For all individuals, the monthly mean $\delta^{18}O_{oto}$ was always lower than that of the previous month, resulting in consistently decreasing trends. There were significant differences in $\delta^{18}O_{oto}$ in all months preceding and following ($p < 0.05$, Mann-Whitney U test). As for the differences in $\delta^{18}O_{oto}$ history among sampling areas, there were no significant differences in the monthly mean $\delta^{18}O_{oto}$ of individuals among regions in all months, except for those off the Noto Peninsula and off the Goto Islands in June ($p = 0.049$, Steel-Dwass test).

### 3.2    Relationship between seawater $\delta^{18}O$ and salinity

The $\delta^{18}O_{water}$ ranged from -1.37 ‰ to +0.41 ‰, and relatively low $\delta^{18}O$ values (< -0.50 ‰) were concentrated in the northern East China Sea. Similar to the $\delta^{18}O_{water}$, salinity that ranged from 27.213 to 34.861 tended to be low in the northern East China Sea. Both the $\delta^{18}O_{water}$ and salinity gradually increased from west to east. There was a significant positive linear relationship between the salinity and $\delta^{18}O_{water}$ ($n$=237, $p < 0.0001$, $r^2 = 0.964$) and the coefficients are determined as $a = 0.23$ and b = $-7.54$ in Equation (2). The slope $a$ and intercept $b$ are close to those reported in previous studies in this area (Fig. S1; Horikawa et al., 2015; Kodaira et al., 2016).

### 3.3    Estimated migration history

Seasonal movements consistent with the locations of spawning grounds and sampling points were successfully illustrated for all individuals by applying systematic adjustments for the habitat depth assumption (Fig. 3, Fig. S2, S3). For all individuals, the estimated distribution areas (i.e., where the difference of $\delta^{18}O_{est}$ and $\delta^{18}O_{oto}$ was smaller than 0.18) were located in the western offshore of Kyushu (south of 34 °N) until mid June to early July from hatching, regardless of the habitat depth assumption (Fig. 4). The estimated distribution areas at the date closest to the time of hatching included areas where eggs were found in surveys for most individuals, regardless of the habitat depth assumption. These areas were shown to be west offshore of Kyushu (Fig. 3, Fig. S3, S3). However, after July, the estimated distribution areas varied significantly with the depth assumptions. In general, the estimated distribution areas calculated at a depth of 10 m moved rapidly northward, whereas those calculated at 30 m depth showed limited movement (Fig. 4). For individuals off the Noto Peninsula, the estimated distribution areas at the time of sampling were located near the actual sampling point when the habitat depth was assumed to be 10 m (Fig. 5). There were significant inconsistencies between the estimated distribution areas and sampling points calculated at 30m (Fig. 5). In contrast, for the Tsushima Strait and the off Goto Islands individuals, the estimated area came close to the sampling point when the depth was assumed to be 30 m, but became far away when 10 m was assumed (Fig. 5).

### 3.4 Comparison of hatching dates, BL and daily growth rate differences

There was a significant difference in the hatching dates between the northward migration group ($63 \pm 32$ days from January 1st; mean $\pm$ 1SD) and the resident group ($43 \pm 30$ days from January 1st) (Table 2, $p < 0.02$, Mann-Whitney U test). The mean BL of the northward group was $136.7 \pm 6.4$ mm and that of the resident group was $118.9 \pm 9.4$ mm. The BL of the northward group was significantly larger than that of the resident group (Table 2, $p < 0.001$, Mann-Whitney U test). In spring (March-May), the mean back-calculated BL of the resident group was larger than that of the northward migration group (Fig. 6). However, from June, the mean back-calculated BL of the northward migration group was higher than that of the resident group. The mean BL at July 1st, which roughly corresponds to the timing when migration patterns started to diverge, was significantly larger in the northward migration group (northward migration: $104.8 \pm 14.4$ mm, resident: $95.1 \pm 14.8$ mm, Mann-Whitney U test, $p < 0.01$, Table 2). When compared by daily age, the normalized deviation of daily growth rate among the three sampling areas was significantly different during 56–147 (with exception for 69, 76-78, and 140) daily age ($p < 0.05$, Kruskal-Wallis test). Between the off Noto Peninsula and off Goto Islands individuals, the post-hoc Steel-Dwass test showed that the normalized deviation of daily growth rate was significantly different during 56–103 (with exception for 69, 76-78, 89, and 97-98) daily age (Fig. 7, $p < 0.05$). Between the off Noto Peninsula and Tsushima Strait individuals, the deviation was significantly different between 71—75, 79–86, 88–139, and 141–147 daily age (Fig. 7, $p < 0.05$) , if the periods that showed significance more than 5 sequential days. When compared by calendar days, there was a significant difference among groups during 113–206 (with exception for 120, 152–157, 170 and 192–193) days from January 1st (late April to late July) ($p < 0.05$, Kruskal-Wallis test). The post-hoc Steel-Dwass test results showed that the normalized deviation of daily growth rate of the Tsushima Strait and off Goto Islands individuals were significantly different from those of the off Noto Peninsula individuals during 114–118 days (late April) and 185–191 (early July) and 194–206 days (middle July to late July) (Fig. 7, $p < 0.05$), respectively, if the periods that showed significance more than 5 sequential days.

## 4    Discussion

In this study, we conducted a high temporal resolution analysis of $\delta^{18}O_{oto}$ for the young-of-the-year Japanese sardines collected in the Sea of Japan in 2015 and 2016 and reconstructed their migration histories based on a simple method that combines $\delta^{18}O_{oto}$ and an ocean data assimilation model. The values of $\delta^{18}O_{oto}$ showed similar declines with daily age, presenting almost no significant differences among individuals that were collected from the northern and southern sampling locations. This was contrary to our hypothesis that higher $\delta^{18}O_{oto}$ value would be associated with fish collected in the northern area due to the cooler sea surface temperature compared to the southern region; this was attributed to the negative correlation between temperature and $\delta^{18}O_{oto}$ (Sakamoto et al., 2017). Moreover, $\delta^{18}O_{water}$ was similar in the northern and southern areas (Fig. S4) and hence, similarity in $\delta^{18}O_{oto}$ profiles in fish collected from the three sampling areas cannot be explained by the compensation of $\delta^{18}O_{water}$. However, these similarities could be attributed to the differences in habitat depth. For the specimens that were collected off the Noto Peninsula, the estimated distribution areas at a depth of 10 m were consistent with the actual sampling areas, but those at a depth of 30 m were significantly different from the actual sampling areas (Fig. 5). In contrast, for the specimens that were collected off Goto Islands and Tsushima Strait, the estimated distribution areas at a depth of 30 m were consistent with the actual sampling areas, whereas those at a depth of 10 m were different (Fig. 5). This suggests that individuals in the northern and southern regions were distributed at different depths in late summer. However, the estimated distribution areas suggested that the northward migration and the resident groups shared the west offshore of Kyushu as a nursery ground, at least until mid-June to early July, regardless of the estimated habitat depth. These results suggest that there were two migration patterns in individuals that hatched offshore of Kyushu in the west: one migration group started to migrate northward using a shallow layer at some period after mid-June, and a resident group that stayed in the southern area of the Sea of Japan at a relatively deeper depth in summer.

The northward migration and the resident groups showed significant differences in growth trajectories. The mean BL in spring (March-May) was larger in the resident group (Fig. 6). This was likely because the mean hatch date of the resident group was earlier than that of the northward migration group (Table 2). No difference in the growth rate was observed between the resident group and the northward migration group up to 50 days after hatching (Fig. 7). However, the northward migration group grew faster 50 days after hatching than the resident group, and they became larger in June, despite their smaller size in spring (March-May). This suggests that individuals that grow relatively well during late spring and early summer within the population migrated northward.

A migration pattern that includes both migratory and resident individuals is called partial migration, which is observed in various animals, including fish (Chapman et al., 2011, 2012). There are three types of seasonal partial migration of fish species (Chapman et al., 2012). The first is non-breeding partial migration, in which migrants and residents breed sympatrically, but spend non-breeding seasons separately. The second type is partial breeding migration, in which migrants and residents share a non-breeding habitat but breed separately. The third is skipped breeding partial migration, in which migrants and residents share a non-breeding habitat, and individuals migrate to breed, but not every year, leading to partial migration. In addition to the above three types of partial migration, a shorter spatio-temporal partial migration is identified: partial diel vertical migration, in which migrants move vertically during the day or night while residents remain at the same depth (more details see Chapman et al., 2012). In the family of Japanese sardine, *Clupea harengus* shows the characteristic of the breeding partial migrants—they share a common feeding ground but migrate to different areas to spawn in the North Sea (Ruzzante et al., 2006). The Celtic Sea populations of *C. harengus* migrate into the Irish Sea for spawning; these migrants grow more slowly than residents and are therefore recruited later to the adult population (Burke et al., 2008). It is often observed that the body sizes of migrants and residents differ in animals who undergo partial migration (e.g., Kerr et al., 2009). Various hypotheses have been proposed regarding the factors that contribute to size differences (Chapman et al., 2011). The traditional "body size" hypothesis states that a larger body size is advantageous for residents because of their high physiological tolerance to adverse winter conditions (Ketterson and Nolan, 1976). However, exceptions to this hypothesis include birds, in which partial migrations have been well studied. Larger individuals of the male great bustard *Otis tarda* have been shown to migrate during the hot summer months; this has been attributed to the low tolerance for higher temperatures of the larger individuals compared to the smaller ones (Alonso et al., 2009). Additionally, larger individuals of tropical kingbirds, which have high energy requirements, tend to migrate (Jahn et al., 2010). Within skipjack tuna (*Katsuwonus pelamis*) tagged and released from the same location in the western North Pacific, higher percentages of larger (46 cm or larger) individuals were collected at higher latitudes than smaller (44 cm or smaller) individuals (Nihira, 1996). In addition, northward-migrated skipjack tuna had higher energy consumption than those at lower latitudes (Aoki et al., 2017). These findings suggest that the higher energy requirements of larger individuals promote migration, even for fish species. Similarly, in the case of the Japanese sardine in the Sea of Japan, larger individuals might migrate to acquire more energy due to the increase in sea surface temperature around Kyushu.

The water temperatures at depths of 10 m and 30 m on the west offshore of Kyushu were almost the same from February to June (Fig. 8). After June, however, the increase in water temperature at 10 m depth accelerated, and there was a difference between the water temperature at the depths of 10 m and 30 m due to the development of stratification. Excessive higher temperatures result in higher energy dissipation and demand substantial energy intake by consumption (Rudstam, 1988; Ito et al., 2013).

The Japanese sardine might have avoided this by selecting one of the following two strategies: horizontally migrating northward at shallower depths or having deeper habitat depths and staying in the same area. Individuals with better growth and larger body size would have higher energy demands (Noguchi et al., 1990). Meanwhile, the abundance of zooplankton in the western Sea of Japan gradually decreases from spring to summer (Hirakawa et al., 1995); and zooplankton abundance and energy content in the northern Sea of Japan are higher than those in the western Sea of Japan during the summer months (Yasuda et al., 2021). The Japanese sardine in the northern offshore area had larger sizes and higher lipid contents (Yasuda et al., 2021). The Japanese sardines that were able to successfully reach the northern offshore area were rewarded with a better prey field. Thus, larger individuals that required more food may have migrated northward to remain in the optimal water temperature range and seek places where food is abundant. As partial migration was accompanied by habitat depth differences, this case may be considered a combination of non-breeding partial migration and partial vertical migration with larger migrants.

Size dependence in the selection of migration patterns provides insights into the mechanisms of population growth. If only larger individuals in summer migrate from the west offshore of Kyushu to reach a better prey field in the northern offshore area, growth rates from late spring to early summer in the southern area may determine the proportion of migratory individuals to the northern offshore area of the Sea of Japan. Enhanced feeding conditions and larval growth in the west offshore of Kyushu could increase the number of young-of-the-years that have better nutritional conditions in the northern offshore, and ultimately increase the reproduction rate of the Tsushima Warm Current stock. Consistent with this hypothesis, historical analysis of the distribution of the Tsushima Warm Current stock showed an abundance increase in the northern offshore area of the Sea of Japan during the high biomass period (Muko et al., 2018). Thus, the west offshore of Kyushu can be considered as a key area for stock fluctuations. However, these discussions have caveats. All of the samples we analyzed $\delta^{18}O_{oto}$ were estimated to have hatched in the west offshore of Kyushu, but the eggs and larvae were found in the wide coastal area of the Sea of Japan, and the contributions of the other areas to the northward migration group have not been determined. Moreover, the proportion of the northward migration group to the recruitment of the Tsushima Warm Current stock remains unclear. Further studies on population structure are needed to understand the importance of the environment west offshore of Kyushu for the stock.

In conclusion, we revealed migrations in the early life stages of the young-of-the-year Japanese sardine in the Sea of Japan in 2015 and 2016. They all hatched and grew west offshore of Kyushu until late June, and may have chosen one of the two strategies to avoid high temperatures within seasonal and geographical limits: migrating northward or moving deeper. Relatively well-grown individuals in summer were more likely to migrate northward to better feeding grounds, thereby suggesting that the environmental conditions in spawning and nursery grounds west offshore of Kyushu may be important for total energy acquisition during the first year of life. In future, it will be necessary to elucidate the factors that control the growth of individuals west of Kyushu. In addition, further analysis of the population structure is necessary to examine the contribution of individuals from spawning grounds in the Sea of Japan to the Tsushima Warm Current Stock.

**Conflict of Interest**

All the authors have no conflict of interest.

**Author Contributions**

TI initiated this project. TI, TY, and MT designed the research plan. MT and TY provided otolith samples. AH conducted a major part of the otolith daily ring measurements. TS designed a method to estimate migration histories and drew the output with TA. TA, TM, and AI performed isotopic analyses of otoliths under the supervision of TI and KN. SK provided water samples and calculated the equation of $\delta^{18}O_{water}$–salinity equation. TI performed isotopic analysis of seawater. TA drafted the original manuscript. TA, TI, TS, and SI extended the analysis and prepared the manuscript. All authors have contributed to the revision of the manuscript.

**Funding**

This study was partly supported by grants from the Fisheries Agency of Japan, Kurita Water and Environment Foundation, and JSPS KAKENHI (16H02944, 18H04921, 19H04247, 21H04735, 21K18653, and 22H05030).

**Acknowledgments**

This is a preprint to be submitted to a journal. After the acceptance of this manuscript, the link to the publication will be made in this preprint. We express special thanks to Dr. Akira Hayashi for providing otolith daily ring analysis data. We thank S. Sakai for technical advice on micromilling and J. Ibuki and S. Namekawa for technical support with the isotopic analysis. T. Setou provided the hydrodynamic model outputs. The authors acknowledge T. Goto, Y. Ishihara, T. Kodama, H. Kurota, and N. Nanjo for their assistance with the seawater sampling.

**Data Availability**

The original contributions presented in the study are publicly available for otolith microchemistry analyses. Those data will be found here after the acceptance of this manuscript: doi: 10.6084/m9.figshare.25241842. Other data supporting the conclusions of this article will be made available by the authors, without undue reservation.

## Tables

TABLE 1. TABLE 1. Summary of sample data showing the sampling date, number of sampling stations and samples as well as $\delta^{18}O$ values.

| Area | Year | Sampling date | Number of sampling stations | Number of samples | |
|------|------|---------------|------------------------------|-------------------|---|
| | | | | Daily rings analysis | $\delta^{18}O$ analysis |
| off Noto Peninsula | 2015 | Aug. 23rd | 2 | 5 | 2 |
| | 2016 | Aug. 25th and 29th | 3 | 13 | 3 |
| Tsushima Strait | 2015 | Sep. 2nd−5th | 6 | 84 | 2 |
| | 2016 | Sep. 5th−8th | 4 | 23 | 3 |
| off Goto Islands | 2015 | Sep. 12th and 13th | 2 | 20 | 4 |
| | 2016 | Sep. 13th | 1 | 9 | 3 |

TABLE 2. Summary of each characteristic compared by Mann-Whitney U test among groups with different migration patterns.

| Comparison | Northward migration group | | Resident group | | $p$ |
|------------|---------------------------|------|----------------|------|-----|
| | Mean | SD | Mean | SD | |
| Hatching dates (day) | 63 | 32 | 43 | 30 | < 0.02 |
| BL (mm) | 136.7 | 6.4 | 118.9 | 9.4 | < 0.001 |
| Back-calculated BL at July 1st (mm) | 104.8 | 14.4 | 95.1 | 14.8 | < 0.01 |

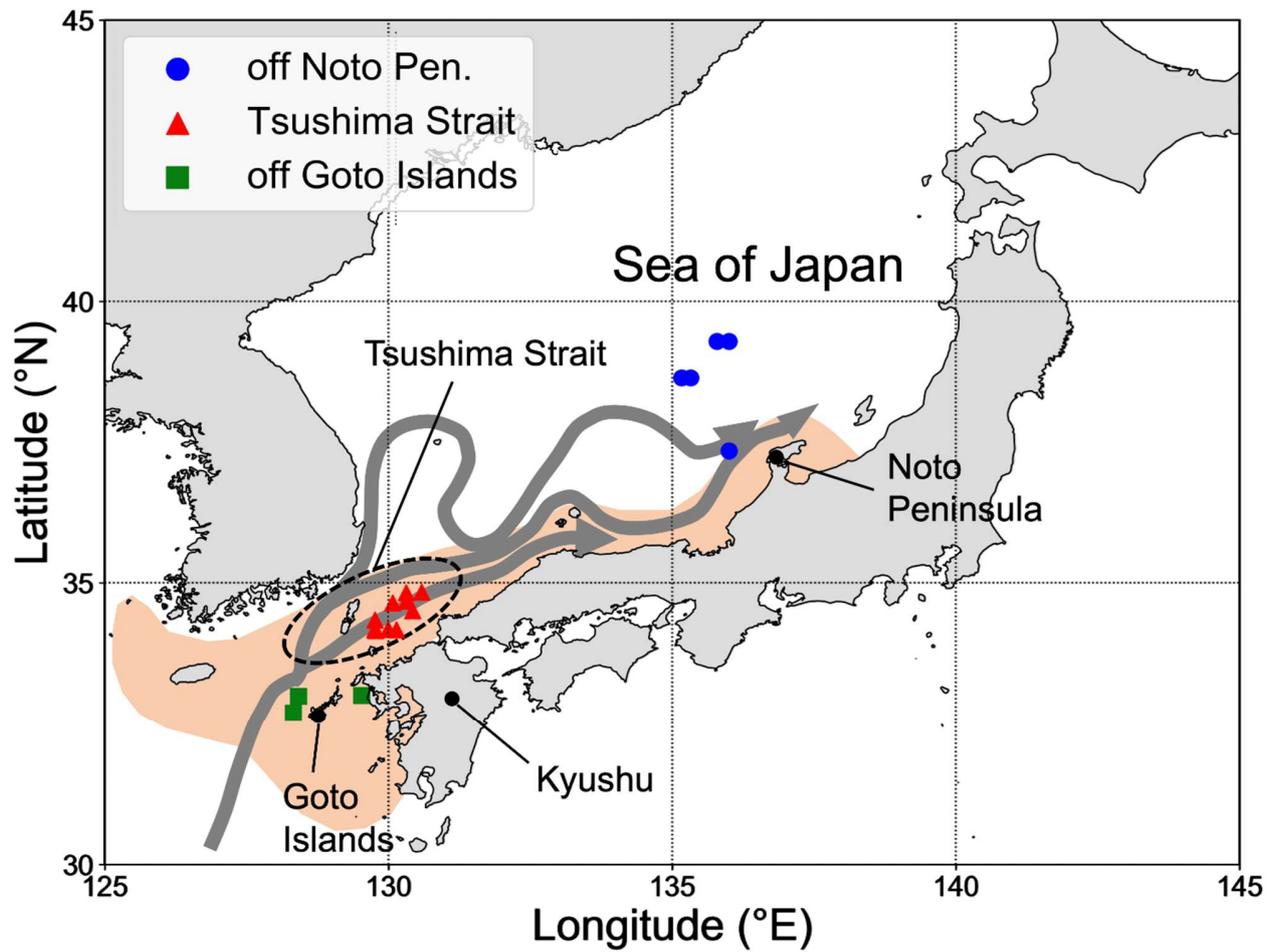

Fig. 1: Study area in the Tsushima Warm Current stock. Blue, red, and green points indicate the locations of sampling stations. Gray arrows show the branches of the Tsushima Warm Current (based on Yabe et al., 2021). Orange area indicates the spawning ground of Japanese sardine in the Sea of Japan (based on Yasuda et al., 2019).

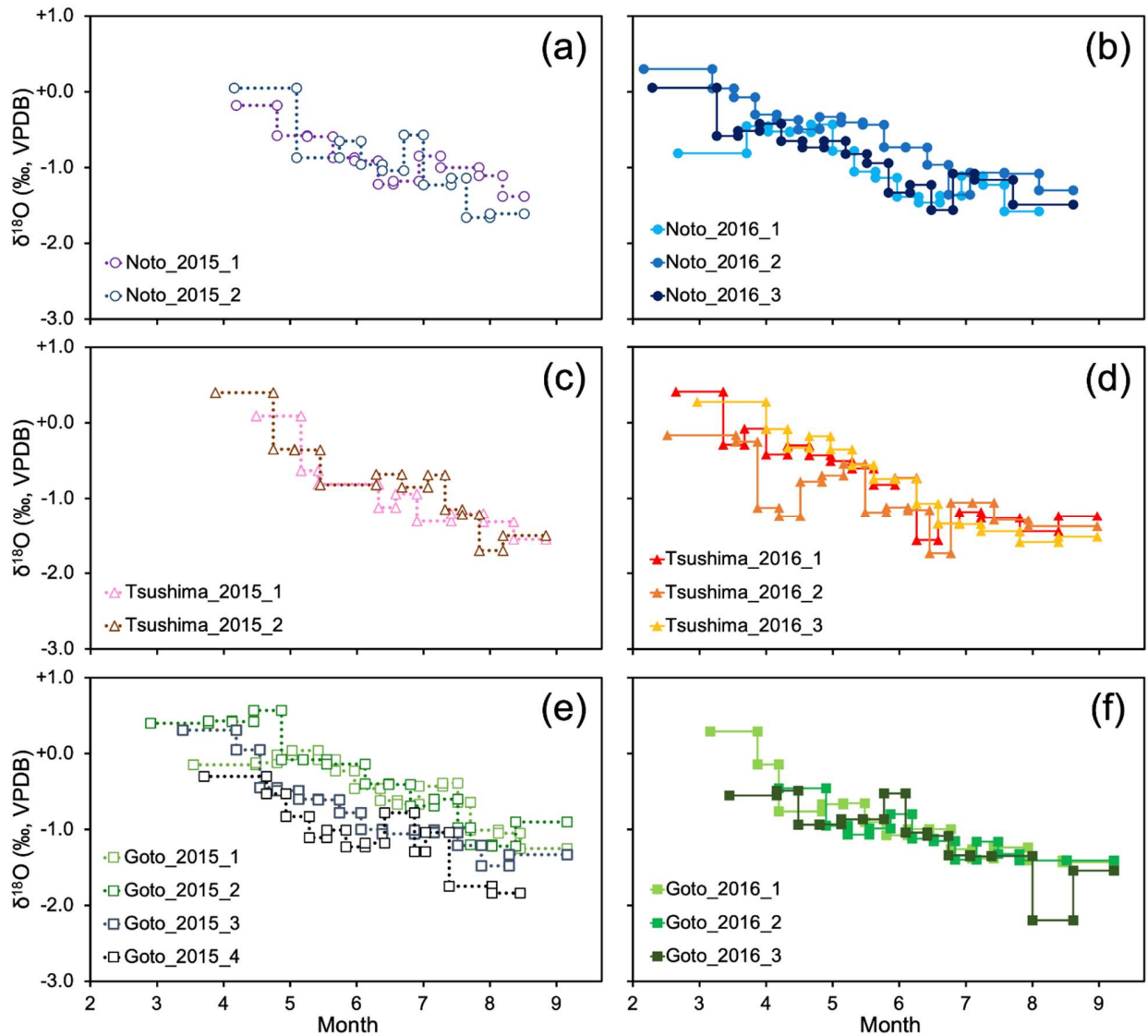

Fig. 2: The $\delta^{18}O_{oto}$ profiles. The left panel shows the $\delta^{18}O_{oto}$ profile of individuals that were collected in 2015 (a, c, e), and the right panel shows the $\delta^{18}O_{oto}$ profile of individuals that were collected in 2016 (b, d, f). The upper column shows the result of off Noto Peninsula (a, b), the middle column shows Tsushima Strait (c, d), and the lower column shows off Goto Islands (e, f).

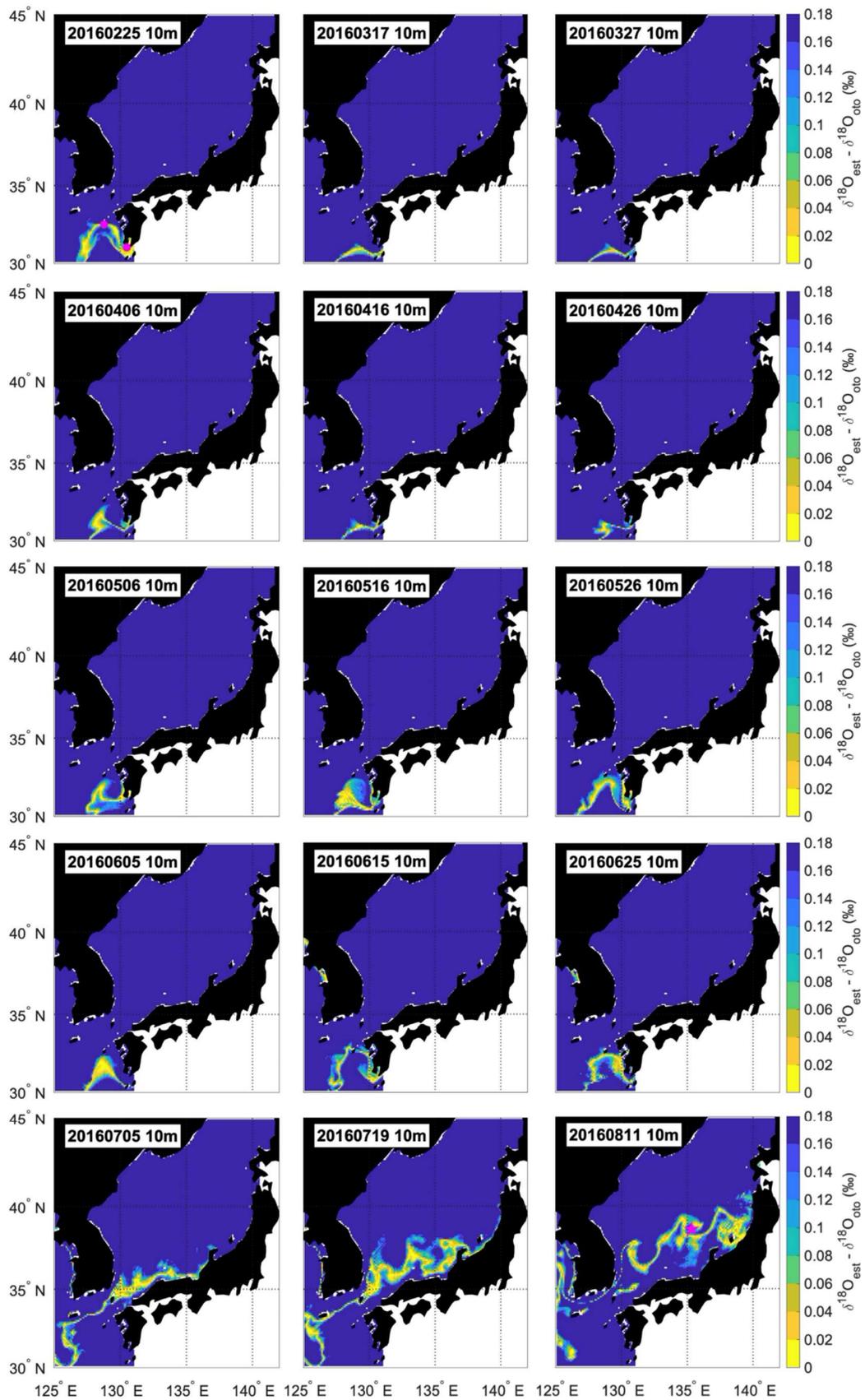

Fig. 3: Example of estimated distribution areas of the individual collected off Noto Peninsula in 2016 (Noto_2016_3), assuming the habitat depth of 10 m. The blue to yellow gradation indicates the estimated distribution areas. Spawning grounds are presented as pink circles. Sampling station is presented as pink star. The upper leftmost figure shows the estimated distribution area of the nearest hatching date and the lower rightmost figure shows the estimated distribution area of the nearest sampling date.

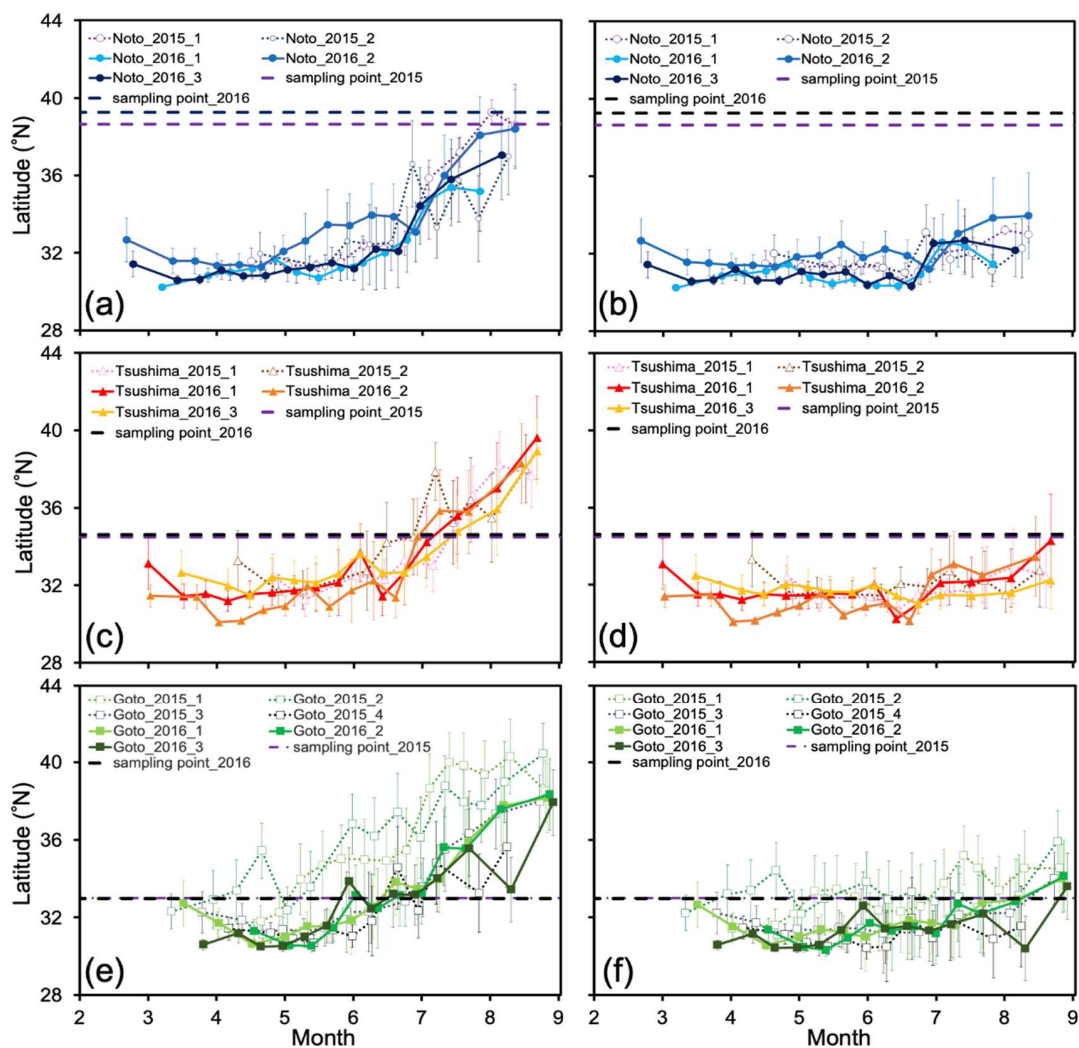

Fig. 4: Temporal variation in mean latitude of estimated distribution areas for each individual. The vertical bars in each plot indicate the standard deviation. The upper panel shows the latitudinal variation of individuals from off Noto Peninsula (a, b), the middle panel shows Tsushima Strait (c, d), and the lower panel shows off Goto Islands (e, f). The left column shows the results of 10 m depth (a, c, e), and the right column shows 30 m depth (b, d, f). The purple and black dashed lines indicate the latitude of the sampling points in 2015 and 2016.

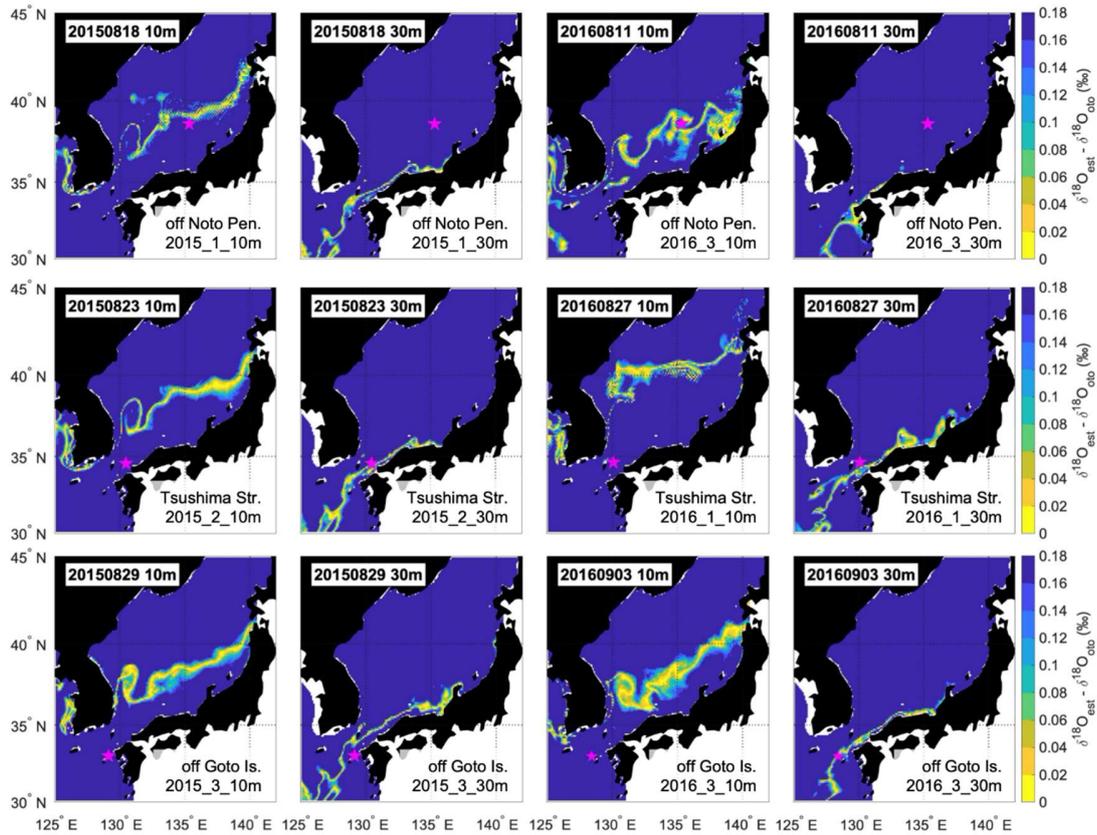

Fig. 5: The estimated distribution areas nearest to the date of sampling, calculated at depths of 10 and 30 m for one representative individual from each sampling area and year. Sample ID and calculated water depth are shown in the lower right corner of each panel. The upper row shows the result of off Noto Peninsula, the middle row shows Tsushima Strait, and the lower row shows off Goto Islands. The left two columns show the estimated distribution areas for individuals sampled in 2016, and the right two columns show the estimated distribution areas for individuals sampled in 2015.

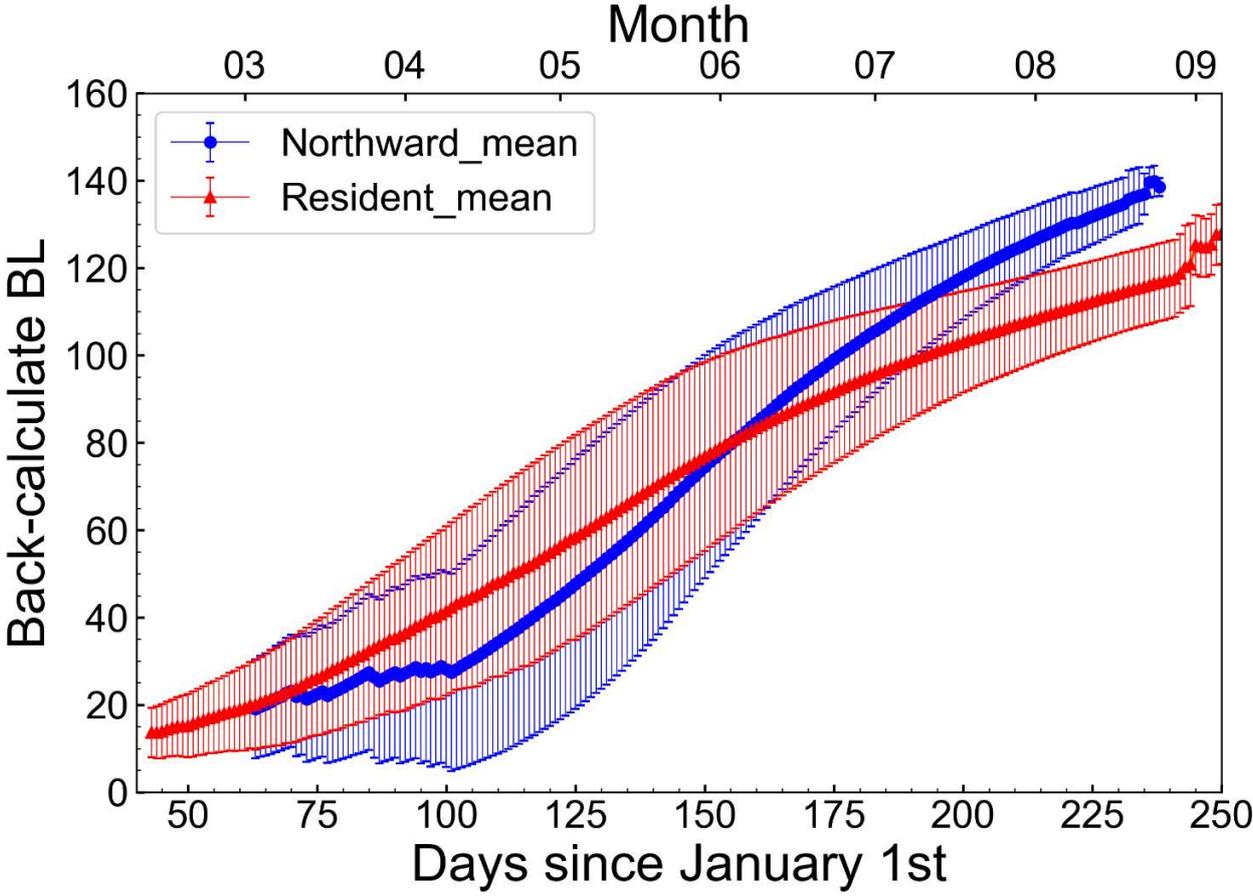

Fig. 6: Back-calculate body length (BL) of the northward migration group (blue line) and the resident group (red line) at each calendar day.

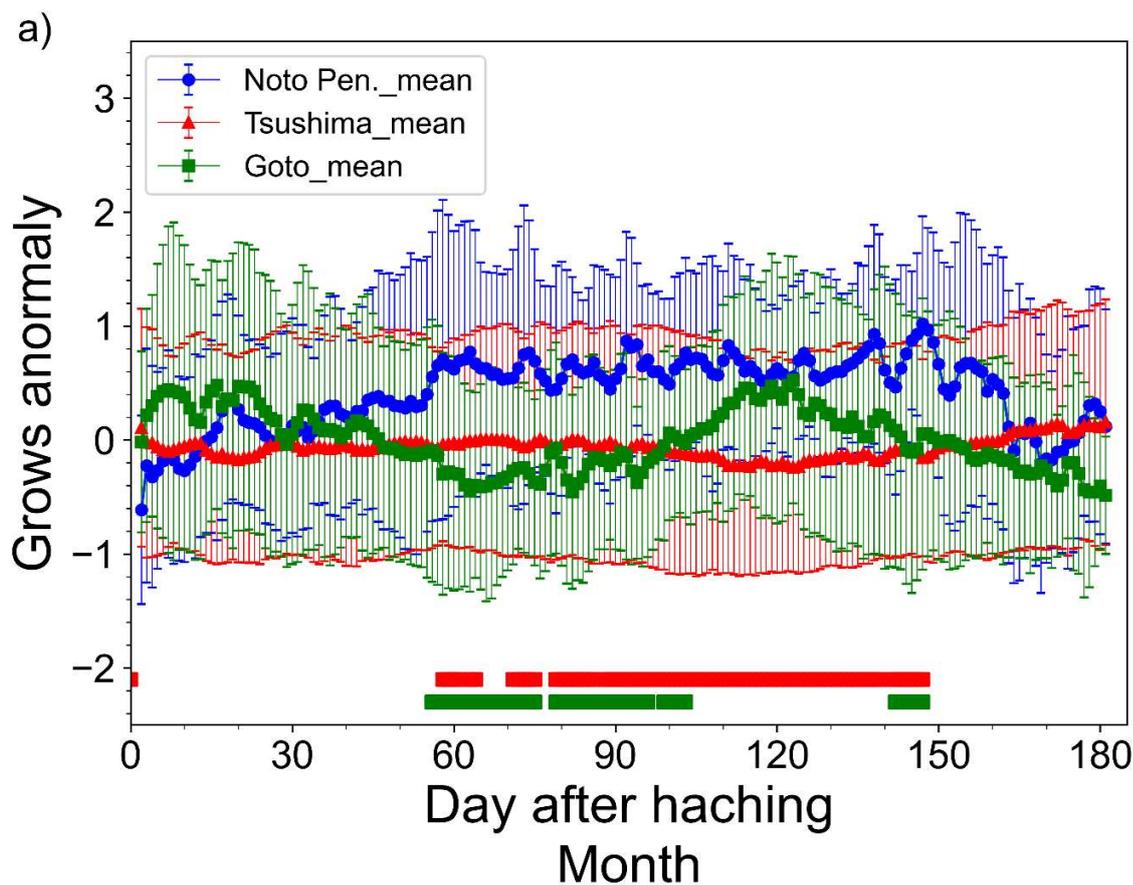

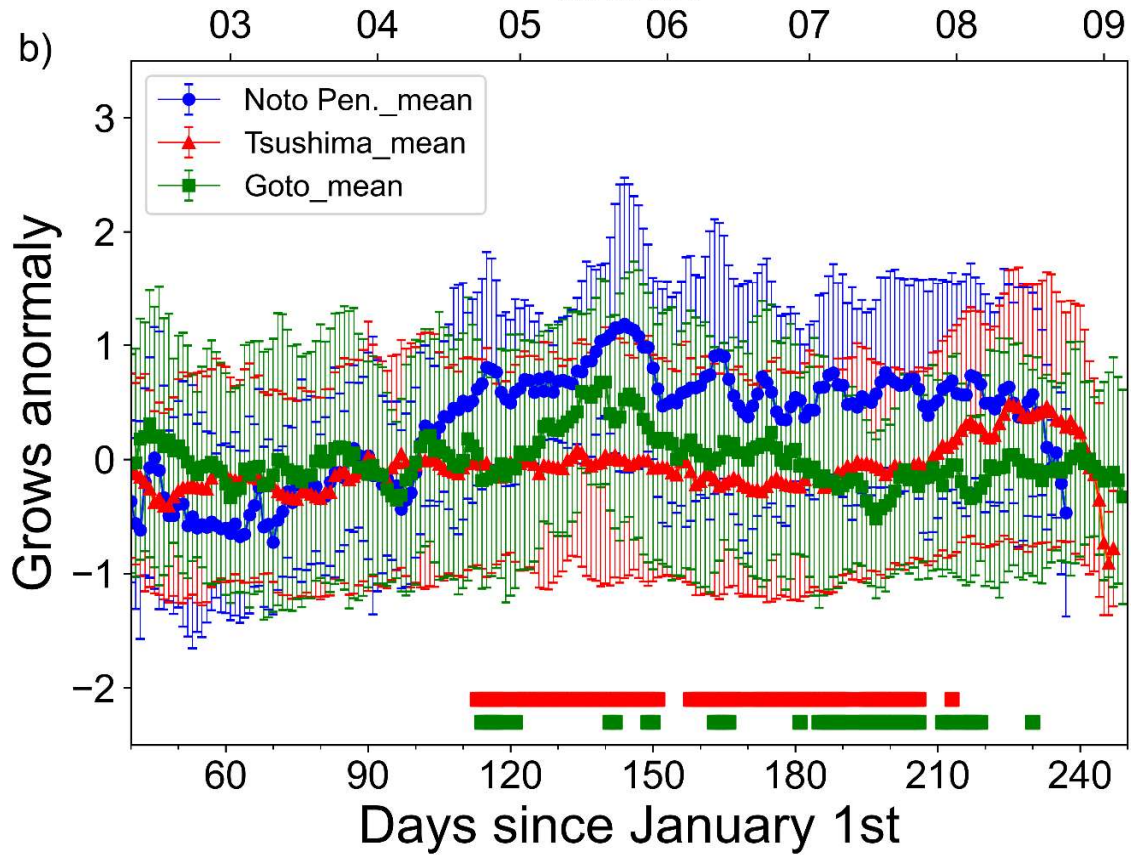

Fig. 7: Normalized deviation of daily growth rate averaged for each sampling areas (blue line: off Noto Peninsula, red line: Tsushima strait, and green line: off Goto Islands). The left panel shows daily age (a) and the right panel shows calendar day (b). The red and green intervals at the bottom of the figure indicate the range of significant differences in the daily growth rate between off Noto Peninsula and Tsushima Strait (red), and between off Noto Peninsula and off Goto Islands (green), respectively ($p <$ 0.05, Steel-Dwass test).

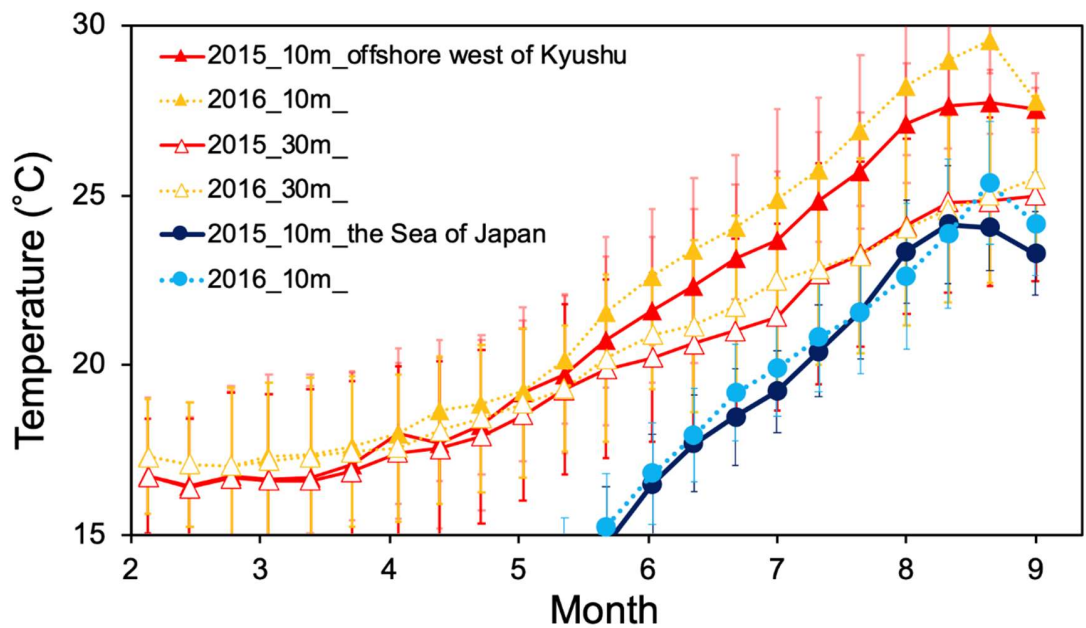

Fig. 8: Water temperature variations extracted from FRA-ROMS on the offshore west of Kyushu (127–130°E, 30–34°N; solid triangles indicate 10 m depth, open triangles indicate 30 m depth) and the northern offshore area of the Sea of Japan (130–140 °E, 36–40 °N; solid circles indicate 10 m depth). The error bar of each plot shows the standard deviation in the area.

# *Supplementary Material*

## 5    Supplementary Tables and Figures

### 5.1    Supplementary Tables

TABLE S1. Summary of analytical results for all individuals analyzed for stable isotope analysis. The data will be available from doi: 10.6084/m9.figshare.25241842 after this manuscript is accepted.

### 5.2    Supplementary Figures

Supplementary video 1. An example of otolith milling process in a one-year-old Japanese sardine collected in the Sea of Japan. The distance from the core to the edge of this otolith was 1428 $\mu$m. This is not the sample used in this study, but is shown here as an example. Double-click on the image to play it.

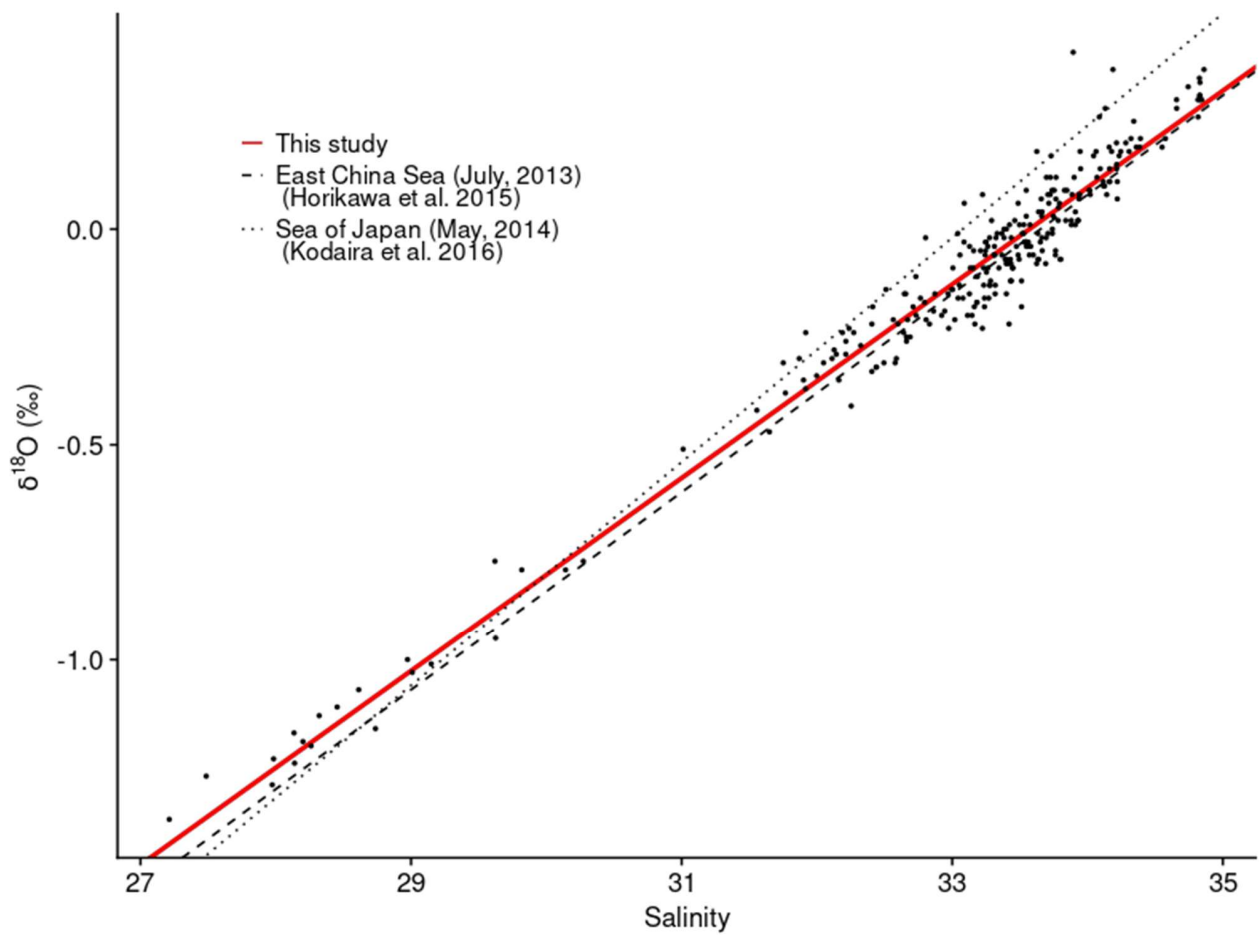

Figures S1. Comparison of $\delta^{18}O_{water}$ and salinity in the Sea of Japan. Filled circles indicate the measured values in this study. The solid red line shows the relation of this study, the dashed black line shows the relation of Horikawa et al. (2015), and the dotted black line shows the relation of kodaira et al. (2016).

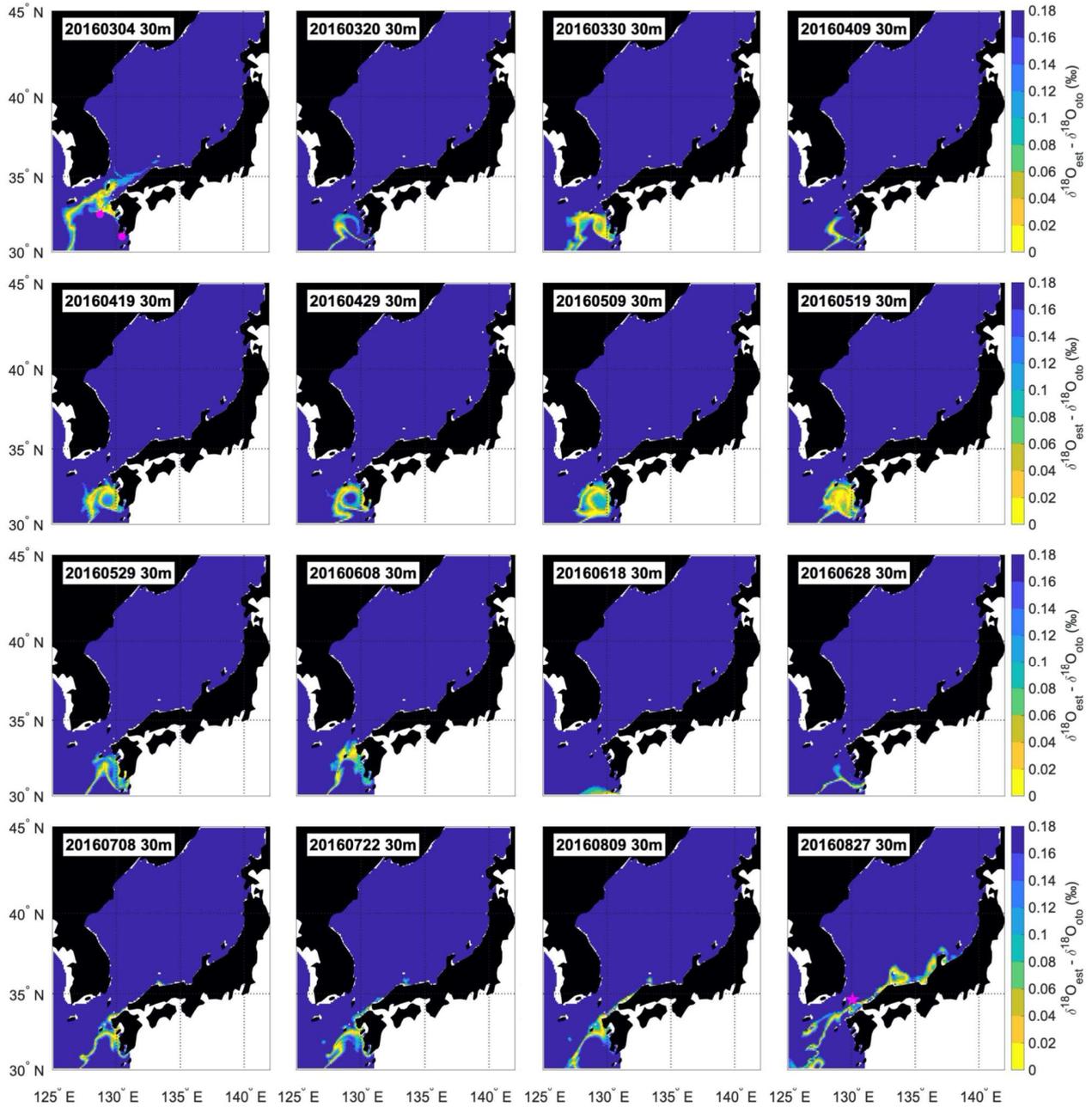

Figures S2. Example of estimated distribution areas of the individual collected in Tsushima Strait in 2016 (Tsushima Strait_2016_1), assuming the distribution depth of 30 m. The blue to yellow gradation indicates the estimated distribution areas. Spawning grounds are presented as pink circles. Sampling area is presented as pink star. The upper leftmost figure shows the estimated distribution area of the nearest hatching date and the lower rightmost figure shows the estimated distribution area of the nearest sampling date.

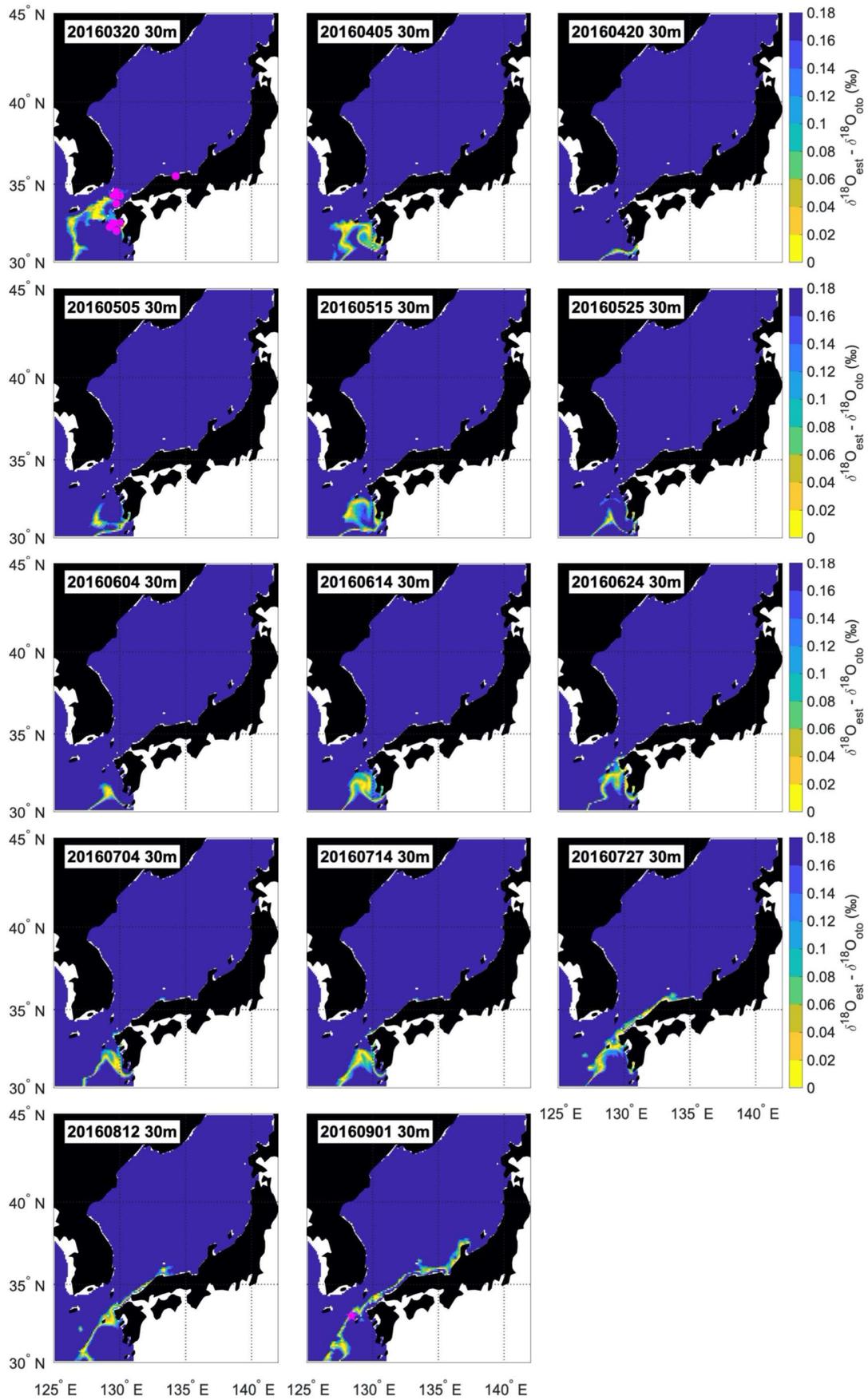

Figures S3. Example of estimated distribution areas of the individual collected off Goto Islands in 2016 (Off Goto Islands_2016_1), assuming the distribution depth of 30 m. The blue to yellow gradation indicates the estimated distribution areas. Spawning grounds are presented as pink circles. Sampling area is presented as pink star. The upper leftmost figure shows the estimated distribution area of the nearest hatching date and the lower rightmost figure shows the estimated distribution area of the nearest sampling date.

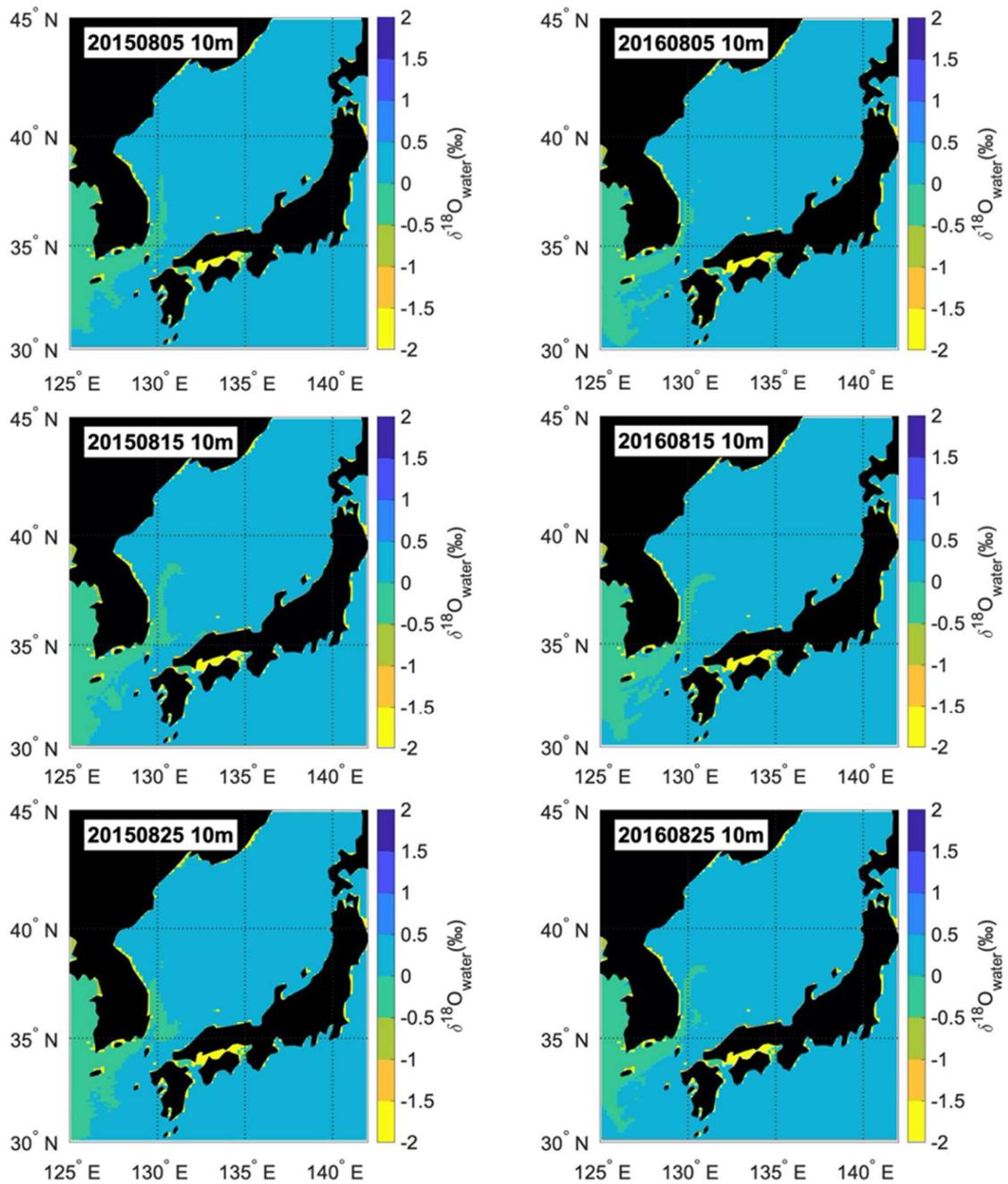

Figure S4. Spatial distribution of $\delta^{18}O_{water}$ in the Sea of Japan at a depth of 10 m in summer. $\delta^{18}O_{water}$ is calculated by substituting the salinity field of FRA-ROMS into equation (2). The date of the calculation is shown in the upper left corner of each panel. The left column shows the results for 2015 and the right column shows the results for 2016.